# Light reflection from a metal surface with subwavelength cavities

Cheng-ping Huang, Jia-qi Li, Qian-jin Wang, Xiao-gang Yin, and Yong-yuan Zhu

*National Laboratory of Solid State Microstructures, Nanjing University*

*Nanjing 210093, P.R. China*

Abstract

The interaction of light with the localized/delocalized system, i.e., a metal surface with rectangular cavities of finite depth, has been studied. Reflection spectrum has been measured in the optical frequencies and resonant minima have been observed. We have developed an analytical model, which agrees well with the experiment. The localized waveguide resonance and delocalized surface resonance have been identified and discussed. The results may be useful for manipulating the coupling between light and matters. [Appl.Phys.Lett. 93, 081917 (2008)]

Email: cphuang@njut.edu.cn, yyzhu@nju.edu.cn

Plasmonic surface has exhibited a series of abilities in manipulating the propagation of light, such as transmission enhancement with the perforated metal films [1, 2], beaming light from an aperture surrounded by the surface corrugations [3, 4], and subwavelength localization using the metallic waveguides [5-7] etc. These effects are generally believed to be linked to the propagating mode, i.e., the surface-plasmon polariton (SPP) mode originating from coupling of light to surface charges. Recently, the metallic systems that simultaneously support propagating and localized modes have attracted much attention [8]. One example of this system is a structure consisting of spherical voids buried closely below a continuous metal surface [9-12]. Due to the localized surface plasmons excited in the spherical voids or SPP resonance on the continuous metal surface, resonant dips can be observed in the reflection spectrum [10]. Moreover, the SPP mode can be further enhanced by coupling to the localized void plasmons, which provides the possibility for developing the SPP-based sensors with high sensitivity [12].

A metallic grating with one-dimensional deep grooves presents another example of such a mixed system [8, 13]. Besides the propagating surface mode, the grating supports in deep grooves the localized waveguide resonance (there is no cutoff for the fundamental groove mode), which play the same role as the void plasmons localized in the spherical cavities. Theory and experiment suggested that, when illuminated with the *p*-polarized light, reflection dips related to these modes will be resulted [13]. In this paper, we show that the similar effect will occur for a metal surface having two-dimensional rectangular cavities of finite depth (the cavity bottoms are closed). Although the rectangular cavity has a cutoff wavelength (beyond which the basic mode is evanescent), the localized resonance can be established in certain wavelength regions, giving rise to the reflection minimum. Nonetheless, because of the cutoff effect of cavity, the minima associated with the localized and delocalized modes have different dependence on the variation of cavity depth. This makes a difference from the case presented in the metallic grooves.

For convenience, we assume that rectangular subwavelength cavities arranged in a square array of lattice constant *d* are cut into the surface of an infinitely thick metal, where the area of cavity is $a \times b$ (*a* in x direction and *b* in z direction) and the cavity depth is *h* (*h* in y direction). The permittivity of metal, interface medium, and filling medium in the cavities are $\varepsilon_m$, $\varepsilon_d$ and $\varepsilon_h$, respectively. A TM-polarized light (the magnetic field is along the z axis) is incident upon the metal surface and reflected into the interface medium, where x-y plane is the incident plane and $\varphi_i$ the incident angle. Figure 1 shows a schematic view of the structure under investigation.

In our experiment, a single-mode optical fiber coupler was used, which has a diameter of fiber center of $9\,\mu m$. We cut one end of the fiber coupler and coated the cross section by sputtering with the gold film of a few nanometers. Then, a square array of dielectric pillars was fabricated into the fiber with the focused-ion-beam system (strata FIB 201, FEI company, 30 keV Ga ions), where the sizes of pillars are the same as those of cavities. Finally, the gap of pillar arrays was filled and further covered with the gold, thus forming the required structures ($\varepsilon_d = \varepsilon_h = 2.1316$). In the measurement, an incident light was coupled into the optical fiber and impinged normally upon the cavity arrays. The reflected signal was split via the coupler and collected finally by an optical-spectrum analyzer (ANDO AQ-6315A).

It should be mentioned that, in the measured spectral range, nonzero-order diffracted waves are evanescent or propagating with a larger deflect angle with respect to the normal of cavity arrays. Thus, they will be confined to metal surface or get away from the fiber, and only the zero-order reflection can be detected. In addition, the reflectance of a flat gold/fiber interface has been used as a reference. Without loss of generality, a sample with the lattice constant $d$=800nm and cavity size $a \times b \times h = 300 \times 300 \times 350\,nm^3$ has been fabricated with the method described above (the area of cavity arrays is $12 \times 12\,\mu m^2$). Figure 2(a) shows the measured zero-order reflection spectrum of the sample (the circles). Compared to the homogeneous reflectance of a flat metal surface (close to unity at the optical frequencies), significant reflection modulation has been introduced into the spectrum. In the spectral range, there are two dominant reflection minima: one locates at the wavelength 963nm and the other is around 1275nm (the weak amplitude oscillation off resonance may be resulted from a slight surface corrugation in the fabricated sample). The observed reflectivity at two minima is about 37% and 53%, respectively. The results suggest that strong optical resonance exists in the cavity arrays.

To understand the optical response of cavity arrays, an analytical model has been developed for the system sketched in Fig. 1. In the theoretical analysis, the diffracted wave fields in the interface medium were expanded in Fourier series; and in the cavities, we assumed that only the fundamental mode can be excited by the incident light. In addition, the surface-impedance boundary condition has been used for the entire metallic interface. Here we summarize some of the results as follows. Firstly, the propagation constant of fundamental cavity mode is written as

$$q_0 = \sqrt{k_h^2 - \alpha^2 - \beta^2}, \tag{1}$$

where $k_h = k_0\sqrt{\varepsilon_h}$ ($k_0$ is the wavevector in free space), $\alpha$ and $\beta$ are determined, respectively, by $\alpha \tan(\alpha a/2) = -ik_0\varepsilon_h/\sqrt{\varepsilon_m}$ and $\beta \tan(\beta b/2) = -ik_0\sqrt{\varepsilon_m}$ [14]. The cutoff wavelength is thus deduced to be $\lambda_c = 2\pi\sqrt{\varepsilon_h/(\alpha^2 + \beta^2)}$. Secondly, the distribution of wave fields is dependent on the function $F(\lambda)$, with

$$F(\lambda)^{-1} = \frac{1+\theta_-}{1-\sigma\varepsilon_m^{1/2}} - \frac{1-\theta_+}{1+\sigma\varepsilon_m^{1/2}} e^{2iq_0 h}. \tag{2}$$

Here $\sigma = k_0 q_0^{-1}(1 - \alpha^2 k_h^{-2})$ and $\theta_\pm$ can be expressed as

$$\theta_\pm = \frac{\sigma \pm \varepsilon_m^{-1/2}}{\sin c(\alpha a/2)} \sum_{n=-\infty}^{+\infty} \frac{w\varepsilon_d g_n s_n}{u_n + \varepsilon_d \varepsilon_m^{-1/2}}, \tag{3}$$

where $w = ab/d^2$ is the normalized area of the cavities, the summation on $n$ involves the diffraction orders above the metal surface, with $g_n = \sin c(k_0 \gamma_n a/2)$, $s_n = (1/a)\int_{-a/2}^{a/2} e^{-ik_0\gamma_n x} \cos(\alpha x)dx$, $u_n = \sqrt{\varepsilon_d - \gamma_n^2}$, and $\gamma_n = \sqrt{\varepsilon_d}\sin\varphi_i + n\lambda/d$. Lastly, with the variables introduced above, the zero-order reflection can be obtained

$$r_0 = |\kappa_0 - \tau_0 F(\lambda)|^2. \tag{4}$$

Here $\kappa_0 = (u_0 - \varepsilon_d\varepsilon_m^{-1/2})/(u_0 + \varepsilon_d\varepsilon_m^{-1/2})$ and $\tau_0$ has the following form

$$\tau_0 = \frac{2w\varepsilon_d g_0 u_0 s_0 (e^{2iq_0 h} - 1)}{\varepsilon_m^{1/2}(u_0 + \varepsilon_d\varepsilon_m^{-1/2})^2 \sin c(\alpha a/2)}. \tag{5}$$

Zero-order reflection spectrum of the sample has been calculated using Eq. (4) and the result is shown in Fig. 2(a) by the thicker line (the frequency-dependent permittivity of gold is chosen from the experimental data [15]). One can see that the theoretical result also presents the reflection modulation and that a good agreement between theory and experiment is resulted. The calculated reflection minima locate at the wavelength 956nm and 1270nm, respectively, which are very close to the experimental values (963nm and 1275nm). In addition, we have also simulated the reflection spectrum with the finite-difference time-domain method [see Fig. 2(a), the thinner line]. The extended agreement concerning the spectrum shape and positions of reflection minima further strengthens the validity of our results. It is worthy of mentioning that the cutoff wavelength of cavity has been calculated to be 1080nm, which is larger than 956nm of one minimum but remarkably smaller than 1270nm of a second minimum. It demonstrates that the reflection minima of the structure can be associated with propagating and evanescent cavity mode.

We have investigated the influence of some physical and geometrical parameters on the reflection spectrum. Figure 2(b) presents the calculated results with different cavity depth: h=400nm (the open squares), 600nm (the open circles), and 800nm (the solid circles). One can see that, with the increase of cavity depth, the minimum below the cutoff wavelength exhibits significant redshift and more minima will be exhibited. Instead, the position and width of minimum above the cutoff do not change with the cavity depth. This behavior is completely different from that of metallic grooves [13], where the depth of groove dominates all the reflection minima (this is not difficult to understand: the groove mode is propagating and penetration length is the groove depth, while the cavity mode is evanescent above the cutoff and penetration length is determined by the propagation constant). On the other hand, when the incident angle or lattice period is altered, the result (not shown here) is just on the contrary: the former minimum changes slightly, whereas the latter strongly.

To identify the origin of reflection minima, we have plotted in Fig. 3 the dependence of $|F(\lambda)|$ (the open circles) and $\text{Re}[\exp(2iq_0h)]$ (the solid circles) on wavelength [see Eqs. (4) and (5); here h=800nm]. Note that the cutoff wavelength of cavity has been marked by an arrow. As shown, when the wavelength is smaller than the cutoff, $|F(\lambda)|$ is trivial and $\text{Re}[\exp(2iq_0h)] \approx \cos(2q_0h)$ dominates the reflection spectrum (here $q_0$ is mainly real). The condition for minimum can be approximated by $\cos(2q_0h) = -1$ or $2q_0h = (2n+1)\pi$, which is dependent on the cavity depth but not on the incident angle (where $n$ is an integer. When $n=0$, the location of minimum is $\lambda_{\min} \approx 4h\lambda_c / \sqrt{16h^2 + \varepsilon_h^{-1}\lambda_c^2}$). That is, the phase difference between light reflected from metal surface and that from cavity bottoms equals an odd multiple of $\pi$. Correspondingly, the reflected light will cancel each other and the energy is concentrated in the cavities. Thus, this type of minimum can be attributed mainly to the localized waveguide resonance associated with the individual cavities. Compared with the exact position of waveguide resonance, however, the reflection minimum exhibits small wavelength shift, as the cavity mode has been dressed by the diffracted surface waves. When the wavelength is larger than the cutoff wavelength, instead, $\text{Re}[\exp(2iq_0 h)] \approx \exp(-2|q_0|h)$ approaches zero and $|F(\lambda)|$ governs the spectrum (here $q_0$ is mainly imaginary, and, a single cavity cannot be resonant). Then, the condition for minimum can be simplified to $1+\theta_- = 0$ [see Eq. (2)], which is dependent on the incident angle but not on the cavity depth. This condition is just similar to that of enhanced light transmission through subwavelength holes [14], showing that they share a common origin. In this case, the evanescent diffraction modes confined to the metal surface will couple strongly to the evanescent cavity modes localized near the cavity openings. Consequently, the light is trapped near the metal surface, giving rise to a suppression of reflection and an enhancement of surface waves (or coupled mode). Thus, this type of minimum can be attributed to the delocalized surface resonance (or nominally the SPP resonance) rather than waveguide resonance.

Here attention should be paid to the role of the plasmonic nature of the metal itself. Since both localized and delocalized resonances are related to the cavity mode, the metallic cavity walls will play an important role. For the metal with a negative permittivity, in Eq. (1) $\alpha$ is imaginary and $\beta$ is real but smaller than $\pi/b$. This means the cavity mode is the plasmon polariton bounded to the cavity walls. Note that, when the metal is perfect conducting, the cavity mode is just simplified to the classical TE$_{01}$ mode with $\alpha = 0$ and $\beta = \pi/b$. If the gold used here is replaced by aluminum which owns a larger plasma frequency, we will have a smaller $|\alpha|$ and a larger $\beta$ as well as a reduced cutoff wavelength. Correspondingly, both types of reflection minima will shift to the shorter wavelengths, according to the resonance conditions. Furthermore, considering of the effect of the metallic film surface, an additional blueshift of the delocalized resonance can be resulted.

The reflection minima demonstrated above are accompanied by enhanced emission or absorption of incident light. For the localized resonance, the resonant cavities will behave as efficient sources of emissions. With the analytical model we have calculated the first-order reflection spectrum (not shown here). It was found that the zero-order reflection minimum corresponds to the maximum of first-order reflection and that much energy will be reemitted from the metal surface. For the delocalized resonance, however, the first-order reflection will not be present. The decrease of reflection means a greatly enhanced light absorption. Moreover, the latter resonance yields a strong amplification of electric field near the cavity openings (the field enhancement linked to waveguide resonance is relatively small). For the structure studied in Fig. 2(a), for example, the light intensity at the center of cavity opening is about 280 times larger than that of incident light (at 1270nm). This strong electric field may be used for the enhancement of molecular fluorescence [16, 17] etc.

In conclusion, the interaction of light with a metal surface having rectangular cavities of finite depth has been studied. Reflection minima have been observed experimentally and an analytical model has been developed, which agrees well each other. Two types of optical resonances have been identified. The localized waveguide resonance exists below the cutoff wavelength with proper cavity depths. However, the delocalized surface resonance occurs above the cutoff and does not change with the cavity depth. The structure thus belongs to the localized/delocalized systems.

This work was supported by the State Key Program for Basic Research of China (Grant Nos. 2004CB619003 and 2006CB921804), by the National Natural Science Foundation of China (NNSFC, Grant No. 10523001). C.P. Huang would also like to acknowledge partial support from the NNSFC under Grant No. 60606020.

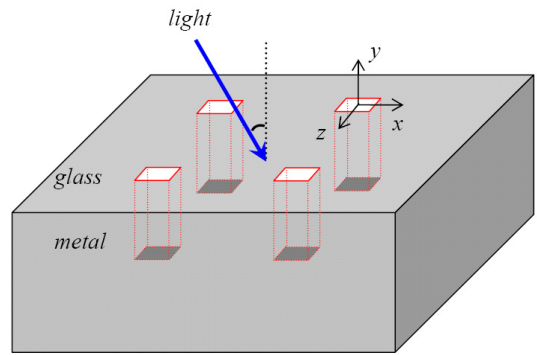

**Fig. 1:** Schematic view of the structure: rectangular cavities arranged in a square array are perforated in the surface of an infinitely thick metal. A TM-polarized light (the magnetic field is in z direction) is incident upon the metal surface and reflected into the interface medium (the cavity bottoms are closed and there is no transmission).

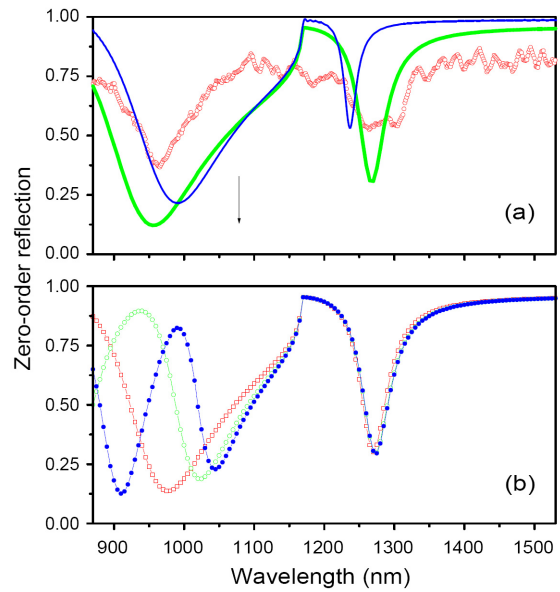

**Fig. 2:** Zero-order reflection of the structure: (a) measured (the circles), analytically calculated (the thicker line) and numerically simulated (the thinner line) spectra with the cavity depth h=350nm; (b) Calculation with different depths h=400nm (the open squares), 600nm (the open circles), and 800nm (the solid circles). Here d=800nm, a=b=300nm, and the permittivity of interface medium and cavity filling medium is 2.1316 (the arrow indicates the cutoff wavelength 1080nm).

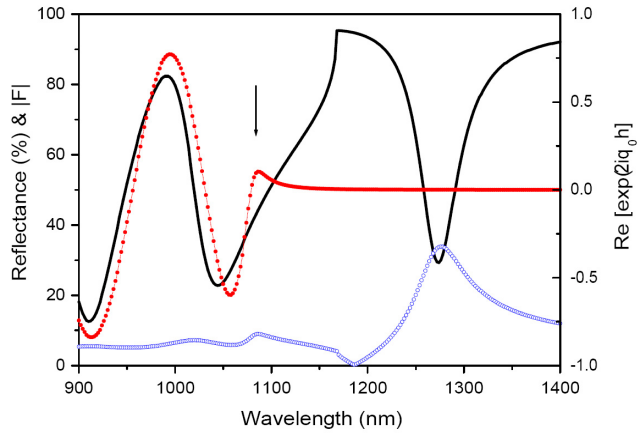

**Fig. 3:** Dependence of $|F(\lambda)|$ (the open circles) and $\text{Re}[\exp(2iq_0h)]$ (the solid circles) on the wavelength, where d=h=800nm, a=b=300nm, and the permittivity of dielectrics is 2.1316. The line represents the zero-order reflection, and the arrow denotes the cutoff wavelength.